\documentclass[letterpaper, 10 pt, conference]{ieeeconf}  
\usepackage{graphicx}
\IEEEoverridecommandlockouts  
\title{\LARGE \bf
Learning a Physical Activity Classifier for a Low-power Embedded Wrist-located Device
}

\author{Ricard Delgado-Gonzalo$^{1}$, Philippe Renevey$^{1}$, Adrian Tarniceriu$^{2}$, Jakub Parak$^{3}$, and Mattia Bertschi$^{1}$
\thanks{$^{1}$R.\,Delgado-Gonzalo, Ph.\,Renevey, and M.\,Bertschi are with the Swiss Center for Electronics and Microtechnology (CSEM), Neuch\^atel, Switzerland; e-mail: {\tt\small ricard.delgado@csem.ch}.}%
\thanks{$^{2}$A.\,Tarniceriu is with PulseOn SA, Neuch\^atel, Switzerland; e-mail : {\tt\small adrian.tarniceriu@pulseon.com}.}%
\thanks{$^{3}$J.\,Parak is with PulseOn Oy, Espoo, Finland; e-mail : {\tt\small jakub.parak@pulseon.com}.}%
}

\begin{document}

\maketitle
\thispagestyle{empty}
\pagestyle{empty}

\begin{abstract}
This article presents and evaluates a novel algorithm for learning a physical activity classifier for a low-power embedded wrist-located device. The overall system is designed for real-time execution and it is implemented in the commercial low-power System-on-Chips nRF51 and nRF52. Results were obtained using a database composed of 140 users containing more than 340~hours of labeled raw acceleration data. The final precision achieved for the most important classes, (Rest, Walk, and Run), was of  96\%, 94\%, and 99\% and it generalizes to compound activities such as XC skiing or Housework. We conclude with a benchmarking of the system in terms of memory footprint and power consumption.
\end{abstract}

\section{Introduction}
Consumer wearable devices are a growing market for monitoring physical activity, sleep, and energy expenditure. These devices are the main promoter of the Quantified Self movement, engaging those who wish to track their own personal data under the premise to improve healthy behaviors. These devices provide feedback to the wearer through device-specific interfaces (\textit{e.g.}, smartphones, web services). Some solutions even provide the means to compare against one's peers or a broader community of users, both of which are useful at increasing overall physical activity through peer-pressure. Originally used by sports and fitness enthusiasts, fitness trackers are now becoming popular as an everyday exercise measurer and motivator for the general public. Step counters can give encouragement to compete with oneself in getting fit and losing weight~\cite{VanWormer2004,LeMasurier2003}.

In the context of human kinetics, wearable devices aim at detecting, classifying, or profiling the kinetic information of the wearer gathered most often through inertial sensors~\cite{Delgado-Gonzalo2017}. These three capabilities are not always present in all systems and are not mutually exclusive either. Among the different inertial sensors, accelerometers have been shown to be the most adapted sensors for a robust recognition of physical activities in wearables systems~\cite{Mathie2004}. Their success and their mainstream usage can be attributed to the fact that they represent a good balance between kinetic information acquired, power consumption, cost, and miniaturization.
In laboratory settings, the most prevalent everyday activities (resting, walking, and running) have been successfully recognized with high precision and recall~\cite{Paerkkae2006,Delgado-Gonzalo2015}. However, one has to be careful at extrapolating these results to out-of-lab monitoring due to the high variability of real-life activities. Direct applicability of the performance results has been challenged in several studies~\cite{Ermes2008}. For example, in~\cite{Delgado-Gonzalo2016} the recognition accuracy of nine patterns decreased from 95.8\% to 66.7\% as the recordings were shifted outside the laboratory.

In the literature, many authors take a principled approach to algorithm design. That is, the activity classification is derived from a carefully hand-picked list of features extracted from the inertial system. These features have a physical meaning that can be exploited by experts in the field to construct a classifier based on well-understood physics. This approach provides good results in protocoled scenarios and in-lab conditions. However, the algorithms based on this approach do not always enjoy the level of generality required for consumer products since they do not take into account the variability in a large user base.

An alternative approach that has been taking steam during the last decade is based on machine learning and data mining. The entry of smartwatches and fitness bands to the consumer market has provided companies with large quantities of data. This data is being used in bulk to iteratively train and improve machine learning algorithms. These algorithms need little or no human intervention and are capable of providing reasonable results in out-of-the-lab conditions. The price that is paid for this automation is the loss of insight on the meaningfulness of the parameters that the algorithm is learning. Moreover, the classes that the system learns may not always correspond to actual different activities due to statistical aberrations derived from the curse of dimensionality.

In the present study, we describe and evaluate a hybrid approach that is data-driven in nature but contains key trainable submodules. The overall system is designed for real-time execution and it is implemented in commercial low-power SoC (nRF51 and nRF52). In the following section, we describe the sensors and protocols that were used to acquire the necessary data to train the physical activity classifier. Then, we proceed with a description of the structure and implementation of the algorithm. Finally, we conclude with a benchmarking of the system in terms of accuracy, memory footprint, and power consumption on both platforms.

\section{Materials}

\subsection{Sensors}
A smart wrist-band integrating a three-axial accelerometer and enough memory to store a full night of raw data was developed by PulseOn\footnote{http://pulseon.com/} and used to acquire raw inertial signals. The sensor integrated in the device is an ST's LIS3DSH and it provides three axial accelerations in the $\pm$8\,g range. Signals are digitized over 12\,bits at a sampling frequency of 25\,Hz.

\subsection{Data acquisition}\label{sec:data}
The acceleration forces from the wrist-located sensors were recorded on 140 individuals (76 male, 64 female) in 18 recording campaigns. The data collection was conducted between 2014 and 2017 in Tampere (Finland), Espoo (Finland), and Neuch\^atel (Switzerland)\footnote{The experimental procedures described in this paper complied with the principles of Helsinki Declaration of 1975, as revised in 2000. All subjects gave informed consent to participate and they had a right to withdraw from the study at any time. Their information was anonymized prior the analysis.} and it was structured in several databases depending on the type of performed activity/protocol:
\begin{itemize}
\item ADLY: Free office work
\item BOUT: Mountain-biking at a variable cadence
\item FVO2: Running outdoors between 30 and 60 minutes at irregular pace
\item LVO2: In-lab timed protocol (40~min) containing sitting, and walking and running on a treadmill at increasing speeds
\item MBDY: Random daily activities such as office work, driving, having lunch, etc.
\item MBOT: Walking, running, and cycling outdoors
\item MDLY: Random daily activities
\item MFOT: Walking, running, cycling, and skiing at comfortable intensity
\item MINT: Random gym activities followed by exercise on a treadmill and cycle ergometer
\item MLAB: In-lab timed protocol (40~min) containing sitting, walking and running on a treadmill, and cycle ergometry
\item MOUT: Running outdoors between 30 and 60 minutes and 4 sets of outdoor cycling
\item MOUTXC: Cross-country skiing
\item MSLP: Overnight sleep sessions
\item MVAL: In-lab timed protocol (40~min) containing sitting, walking and running on a treadmill, and cycle ergometry
\item SDLY: Random daily activities including office work and housekeeping
\item SLAB: In-lab timed protocol (40~min) containing sitting, walking and running on a treadmill, cycle ergometry, and push-ups
\item SPRT: Indoor walking, running, cycle ergometry
\item SPTEST: Indoor walking, running, cycle ergometry, and office work
\end{itemize}

A comprehensive summary of the content of each database is shown in Table~\ref{tb:databases}. A total number of 418 recordings spanning more than 440 hours of raw data was gathered. It is important to notice that some of the 140 test subjects participated in more than one database.

\begin{center}
\begin{table}{
\resizebox{\columnwidth}{!}{%
\hfill{}
\begin{tabular}{c|ccccc}
\hline\hline
Set & Recordings & Duration & Subjects & Age \\ 
 ID& (\#)              & (h) & (\#male/\#female) & (yrs) \\ 
\hline
MOUT & 33 & 21.32 &13m/2f & 33.5$\pm$10.3\\
MVAL & 21 & 13.49 & 11m/10f & 28.3$\pm$5.69\\
MLAB & 24 & 16 & 17m/7f & 24.91$\pm$3.09\\
FVO2 & 24 & 18 & 13m/11f & 36.1$\pm$8.0\\
LVO2 & 24 & 16 & 13m/11f & 36.1$\pm$8.0\\
MSLP & 15 & 79.46 & 13m/2f & 35.9$\pm$10.3\\
MDLY & 12 & 17 & 3m/0f & 32.3$\pm$9.1\\
MINT & 9 & 5.99 & 5m/2f & 35.57$\pm$11.13\\
SDLY & 59 & 67.7 & 15m/13f & 27.21$\pm$6.77\\
SLAB & 47 & 42.9 & 21m/20f & 26.40$\pm$3.27\\
SPRT & 8 & 3.52 & 4m/1f & 32.20$\pm$6.14\\
BOUT & 11 & 6.97 & 6m/1f & 28.66$\pm$6.34\\
ADLY & 18 & 21.9 & 3m/0f & 29.66$\pm$2.08\\
SPTEST & 40 & 18 & 5m/5f & 46.20$\pm$11.81\\
MBDY & 13 & 25.15 & 5m/1f & 30.60$\pm$9.73\\
MBOT & 16 & 13.6 & 5m/0f & 41.25$\pm$15.80\\
MFOT & 15 & 8.7 & 4m/3f & 26.50$\pm$4.23\\
MOUTXC & 29 & 44.87 & 6m/3f & 39.33$\pm$14.35\\ \hline
Total & 418 & 440.57 & 76m/64f & 29.4$\pm$8.58 \\
\hline\hline
\end{tabular}}
\hfill{}
\caption{\label{tb:databases}Summary of the content of each database.}}
\end{table}
\end{center}

Several experts annotated the data choosing among one of the following labels: \textit{Rest}, \textit{Walk}, \textit{Run}, \textit{Bike}, \textit{Office}, \textit{XC skiing}, \textit{Gym}, \textit{Housework}. After consolidation, 340.5~hours from the total of 440.57~hours were annotated. The remaining intervals corresponded to unclear activities or fuzzy transition zones.

\subsection{Algorithm structure}\label{sec:structure}
The algorithm takes as input the raw accelerometer signals at 25\,Hz and outputs the most likely activity among \textit{Rest}, \textit{Walk}, \textit{Run}, \textit{Bike}, or \textit{Other}. The structure is iterative and operates on a sample-by-sample basis, meaning that every new sample from the accelerometer produces a new estimate of the most likely undergoing activity.

\begin{figure}[h]
    \centering
    \includegraphics[width=0.35\columnwidth]{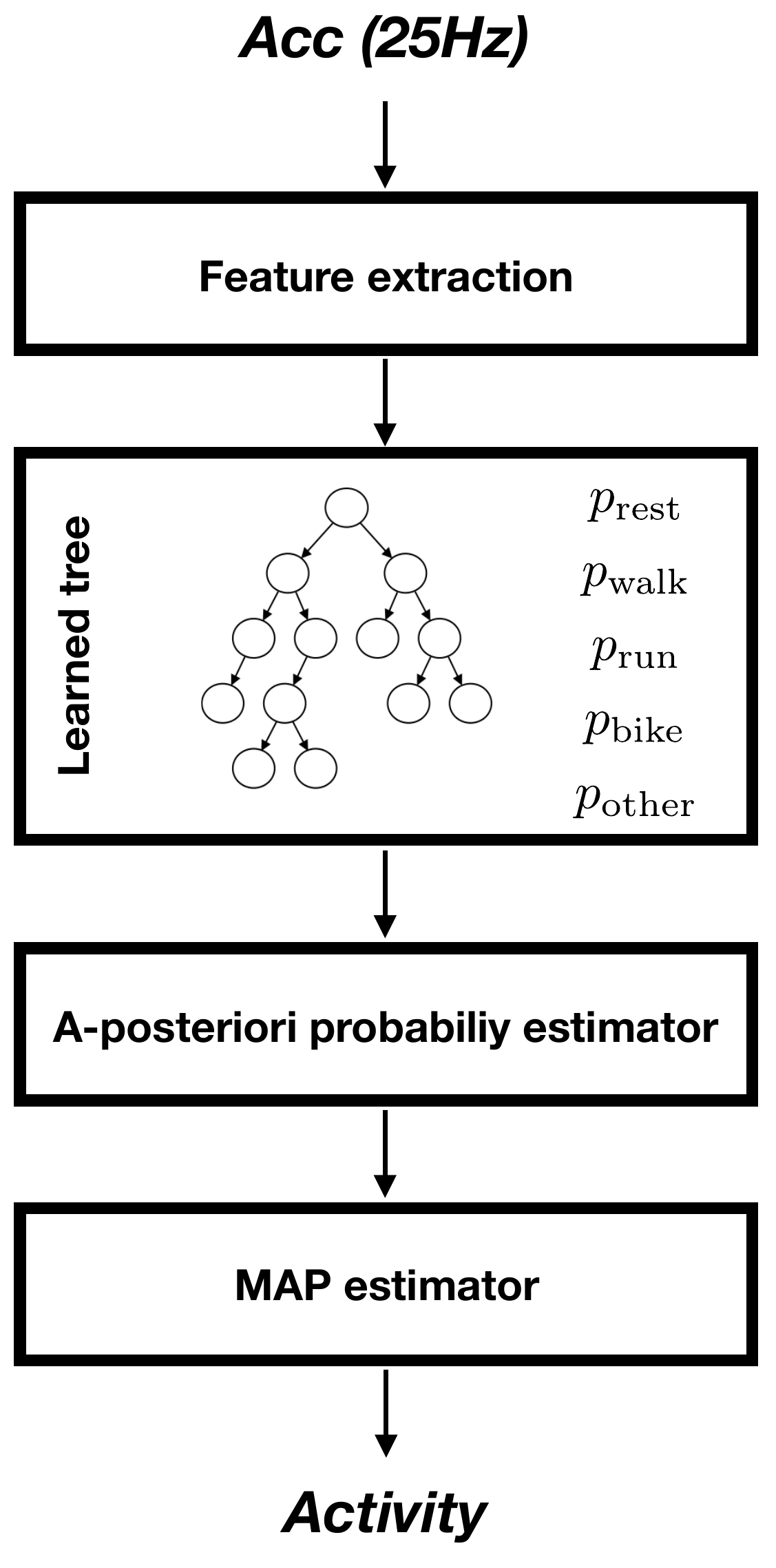}
    \caption{\label{fig:algo}Schematic structure of the embedded algorithm.}
\end{figure}

Internally, the algorithm is composed of four clearly separated parts (see Figure~\ref{fig:algo}). In the first part, several features including signal power, rhythmicity, and frequency stability are extracted from the accelerometer signals. These features are used as predictors in a binary classification tree of depth seven in the second part. Each node of the classification tree contains a different likelihood for each activity. Then, a filter-bank of autoregressive filters of first order (one filter for each activity) is applied independently on each activity. These filters keep temporal consistency across time and their output operates as the set of a-posteriori probabilities for each activity. Finally, the activity with highest probability is selected as output provided that the probability is above a certain threshold; otherwise, \textit{Other} is selected as output.

\subsection{Learning classification graph}
The proposed physical activity classifier was trained with a subset of the recordings described in Section~\ref{sec:data}. More precisely, 31 recordings were randomly selected, and contained multiple activities: outdoor walking and running, road-biking, mountain-biking, indoor cycling, treadmill walking and running, and sleeping. The clear parts of the recordings were manually annotated by several experts following record logs resulting in a total of 1386411 labelled samples ($>$18~hours). Among them, 29.1\% were labeled as \textit{Rest}, 30\% were labeled as \textit{Walk}, 17.5\% were labeled as \textit{Run}, 23\% were labeled as \textit{Bike}, and  0.5\% were labeled as \textit{Other}. The corresponding extracted features were used as predictors to train the binary classification tree using the Gini's diversity index as splitting criterion.

\subsection{Embedded implementation}
The classification algorithm was implemented for two different embedded platforms: the Nordic Semiconductor's nRF52832 and the nRF51822 from the same manufacturer. Each implementation corresponds to a different level of abstraction in nowadays commercial embedded systems and require dedicated instructions in order to take advantage of each platform's strengths and deal with its limitations. The Nordic Semiconductor's nRF52832 SoC incorporates a microprocessor ARM$^{\tiny{\textregistered}}$ Cortex$^{\tiny{\textregistered}}$-M4F. This SoC performs 32-bit integer arithmetic and includes a floating point unit (FPU) with single precision and IEEE 754 compliant. On the other side, the Nordic Semiconductor's nRF51822 SoC incorporates a microprocessor ARM$^{\tiny{\textregistered}}$ Cortex$^{\tiny{\textregistered}}$-M0. This SoC performs a restricted 32-bit integer arithmetic (excluding 32-bit divisions) and does not include an FPU.

The implementation on the nRF52832 was performed in C using the cmsis library\footnote{https://developer.arm.com/embedded/cmsis} for mathematical operations and following a restricted set of the MISRA-C:2012 guidelines. Likewise, the implementation on the nRF51822 was performed in C using fixed-point arithmetic avoiding, when possible, 32-bit divisions and following a restricted set of the MISRA-C:2012 guidelines.

\section{Results and discussion}
\begin{figure}[h]
    \centering
    \includegraphics[width=0.85\columnwidth]{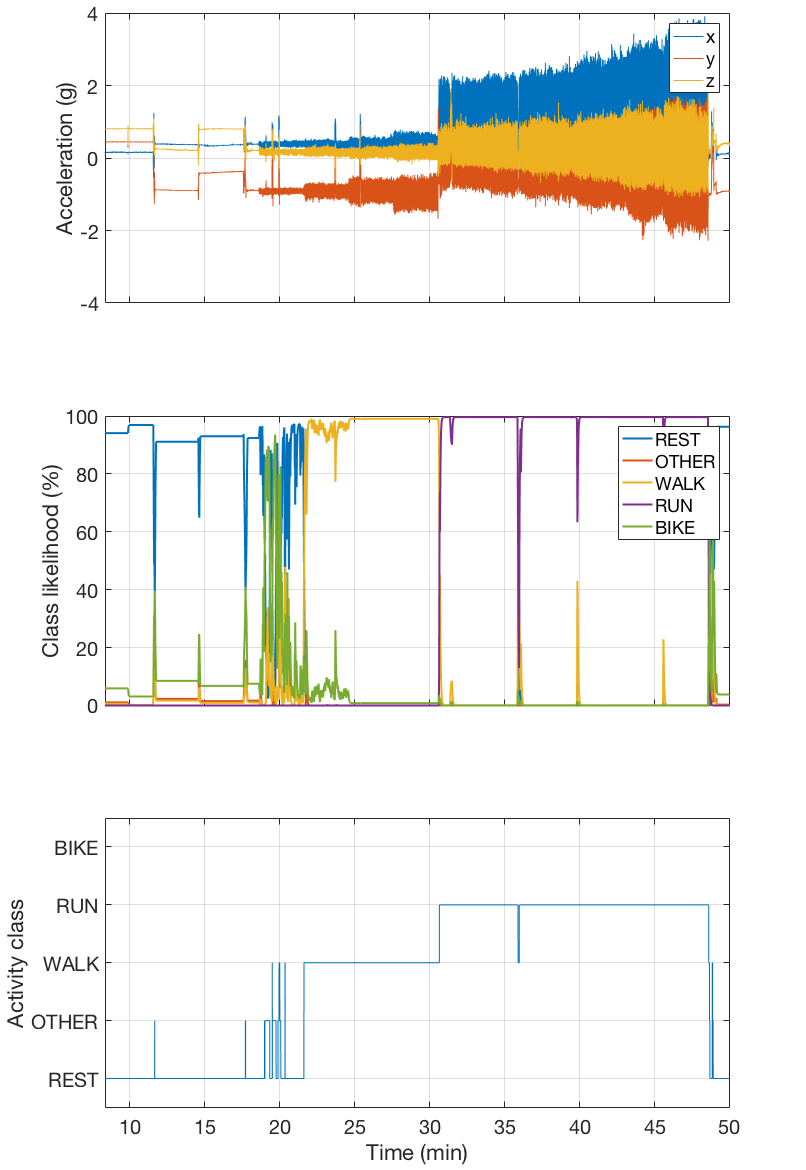}
    \caption{\label{fig:example}Example of the intermediate steps of the algorithm through a dataset from LVO2. (top) raw acceleration signals, (mid) instantaneous class probabilities, (bottom) class estimate.}
\end{figure}

The behavior of the algorithm is best illustrated through an example. In Figure~\ref{fig:example}, we show the process from the raw acceleration signals to a real-time estimation of the physical, going through the instantaneous likelihood of each activity class. This particular example is extracted from the database LVO2 defined in Section~\ref{sec:data}. The raw data is shown in the topmost subfigure, where the three stages of the protocol can clearly be seen: resting, walking, and running at an increasing speed. Then, the a-posteriori probabilities are shown for each class. It is worth to notice that the a-posteriori probabilities ramp up or down following an exponential curve defined by the autoregressive filters of first order discussed in Section~\ref{sec:structure}. Finally, in the bottommost subfigure, the final estimate of the probability is shown. In this example, most of the segments have one clearly dominant class, except the initial moments in the transition between rest and slow walking, where several activities compete to be the most likely.  

\subsection{Accuracy}
In order to evaluate the accuracy of the system, a total of 340.5\,hours were labeled from the databases. In Table~\ref{tb:accuracy}, we show the normalized confusion matrix for the labeled data. The columns represent the estimated class from the classifier and the rows represent the actual labelled classes. In bold we have marked the classes that there is one-to-one correspondence between the estimated and actual classes.

\begin{center}
\begin{table}[h]{
\resizebox{\columnwidth}{!}{%
\hfill{}
\begin{tabular}{c|ccccc|c}
\hline\hline
&\textit{Rest}&\textit{Other}&\textit{Walk}&\textit{Run}&\textit{Bike}&Duration\\
& (\%) & (\%) & (\%) & (\%) & (\%) & (hours) \\
\hline
Resting&\textbf{96.3}&2.80&0.90&0.00&0.16&86.3\\
Walking&1.51&2.68&\textbf{94.66}&0.61&0.53&47.9\\
Running&0.05&0.15&0.91&\textbf{98.88}&0.00&61.8\\
Biking&27.63&25.31&24.74&0.02&\textbf{22.29}&27.4\\
Office working&81.18&13.38&4.29&0.18&0.96&24.3\\
XC skiing&0.60&24.56&53.69&18.74&2.40&33.3\\
Gym&8.25&13.08&53.57&16.6&8.47&2.90\\
Housework&37.82&29.66&25.24&0.05&7.20&56.5\\
\hline\hline
\end{tabular}}
\hfill{}
\caption{\label{tb:accuracy}Classification accuracy of the learned activity classifier.}}
\end{table}
\end{center}

The three main classes, \textit{Rest}, \textit{Walk}, and \textit{Run}, have great classification accuracy and recall ranging from 94.7\% to 98.9\% correctly classified samples. However, biking is equally distributed between \textit{Rest}, \textit{Other}, \textit{Walk}, \textit{Bike}. This stems from the fact that there exist several styles of biking that the classifier did not take into account. For instance, road-biking without pedaling is usually classified as \textit{Rest}, and mountain-biking where the user is not seated is usually classified as \textit{Walk}. This indicates that biking is a multimodal exercise and it would be worthwhile splitting it into more coherent sub-styles.

On the other side, it is interesting to see how the classifier generalizes to activities it has not been trained for. Office working mainly consists on \textit{Rest} and a bit of \textit{Other} (mainly while typing on a keyboard); XC skiing is a combination of \textit{Walk} and \textit{Other}; Gym is mainly \textit{Walk} followed by \textit{Run} and \textit{Other}; and finally, Housework is equally distributed between \textit{Rest}, \textit{Other}, and \textit{Walk}. 

\subsection{Computational load}
For the purpose of measuring the computational load of the different implementations of the algorithm, a standard dataset was defined. This dataset contained 1~minute of raw accelerometer data and was fed to the algorithm cyclically 1000 times. The average execution time was then averaged across all iterations in order to obtain the average execution time per acceleration sample. The test dataset was composed of 58\% of resting, 12.5\% of walking, 4.2\% of running, and 25.3\% of other. For the SoC nRF52832 and nRF51832, the whole library with O3 optimizations with gcc (GNU toolchain from ARM Cortex-M and Cortex-R processors) version 6.0 using Nordic's SDK 12.1. The execution time was then converted to average drained current by using the data contained in the respective SoC's datasheets. A summary of the computational complexity and current consumption is detailed in Table~\ref{tb:benchmark}.

\begin{center}
\begin{table}[h]{
\hfill{}
\begin{tabular}{c|cccccc}
\hline\hline
Platform & Execution time & Current & Complexity\\
& (ms) & (uA) & \\ 
\hline
nRF52832 & 0.067 & 15.3 & 107 KFLOPS\\
nRF51832 & 0.458 & 50.4 & 183 KIPS\\
\hline\hline
\end{tabular}
\hfill{}
\caption{\label{tb:benchmark}Computational load on each platform.}}
\end{table}
\end{center}

\subsection{Memory footprint}
The memory requirements for the nRF52832 and nRF51832 were measured by compiling the whole library with O3 optimizations with gcc (GNU toolchain from ARM Cortex-M and Cortex-R processors) version 6.0 using Nordic's SDK 12.1. The listing (MAP) file produced by the linker was used to generate a complete list of all variables, constants, and functions used.  A summary of the memory requirements is detailed in Table~\ref{tb:memory}.

\begin{center}
\begin{table}[h]{
\hfill{}
\begin{tabular}{c|ccc}
\hline\hline
Platform & RAM & Flash\\ 
 & (Kbytes) & (Kbytes)\\ 
\hline
nRF52832 & 2 & 7.9\\
nRF51832 & 2 & 8.6\\
\hline\hline
\end{tabular}
\hfill{}
\caption{\label{tb:memory}Memory footprint of each implementation of the algorithm.}}
\end{table}
\end{center}

These numbers reflect the small memory footprint consumption of the algorithms. An interesting phenomenon that can be observed in Table~\ref{tb:memory} is that the size of the code is smaller in the nRF52832 than the nRF51832. That is due to the fact that when an FPU is present, the algorithm can be written in a more compact way than the in fixed-point arithmetic, generating less low-level instructions.

\section{Conclusion}
Based on the presented results, we conclude that the proposed approach for learning an activity classifier is capable of generating an algorithm that can be integrated in a real-time embedded system while keeping a high precision and recall for the most common activities (\textit{Rest}, \textit{Walk}, and \textit{Run}). The algorithm generalizes properly to other sports such as XC skiing or Gym, and daily activities such as Office working or Housework. The power consumption and the memory needed represent only a minimal fraction of the overall firmware, and opens the door to a future ultra-low power ASIC implementation.

\addtolength{\textheight}{-12cm}   

\bibliography{references}

\end{document}